\documentclass[preprint,12pt]{aastex}
 
\usepackage{epsfig}
\usepackage{graphicx}

\def\gx{GX~339$-$4}
\def\cx{Cyg~X$-$1}

\def\grs{GRS~1915$+$105}

\def\gs{GS~2023$+$338}

\def\1e{1E~1740.7$-$2942}

\def\xte{XTE~J1550$-$564}
\def\xt{XTE~J1650$-$500}

\newcommand{\sm}      {\mbox{$\rm\,M_{\mathord\odot}$}}         

\shorttitle{Radio emission from \xt}
\shortauthors{Corbel et al.}

\begin{document}

\title{On the Origin of Radio Emission in the X-ray States of \xt\ during the 2001-2002 Outburst.}

\author{S. Corbel\altaffilmark{1}, R.P. Fender\altaffilmark{2}, J.A. Tomsick\altaffilmark{3}, 
A.K. Tzioumis\altaffilmark{4}, S. Tingay\altaffilmark{5}}

\altaffiltext{1}{Universit\'e Paris 7 Denis Diderot and Service d'Astrophysique, UMR AIM, 
CEA Saclay, F-91191 Gif sur Yvette, France.}
\altaffiltext{2}{Astronomical Institute `Anton Pannekoek', University of Amsterdam,
and Center for High Energy Astrophysics, Kruislaan 403, 1098 SJ Amsterdam,
The Netherlands.}       
\altaffiltext{3}{Center for Astrophysics and Space Sciences, University of California  at San Diego, MS 0424, La Jolla,
CA92093, USA.}
\altaffiltext{4}{Australia Telescope National Facility, CSIRO, P.O. Box 76, Epping NSW 1710, Australia. }       
\altaffiltext{5}{Center for Astrophysics and Supercomputing, Swinburne University of Technology, Mail Number 31, P.O. Box 218, Hawthorn, VIC 3122, Australia. }

\begin{abstract}

We report on simultaneous radio and X-ray observations of the black hole candidate \xt\ during the 
course of its 2001-2002 outburst. The scheduling of the observations allowed us to sample the properties
of \xt\ in different X-ray spectral states, namely the hard state, the steep power-law 
state and the thermal dominant state, according to the recent spectral classification of McClintock \& 
Remillard. The hard state is consistent with a compact jet dominating the spectral energy 
distribution at radio frequencies; however, the current data suggest that its contribution 
as direct synchrotron emission at higher energies may not be significant. In that case, \xt\ may 
be dominated by Compton processes (either inverse Comptonization of thermal disk photons and/or SSC from
the base of the compact jet) in the X-ray regime. We, surprisingly, detect a faint level of radio emission in the 
thermal dominant state that may be consistent with the emission of previously ejected material 
interacting with the interstellar medium, similar (but on a smaller angular scale) to what was observed in \xte\ 
by Corbel and co-workers. Based on the properties of radio emission in the steep power-law state of \xt , and 
taking into account the behavior of other black hole candidates (namely GX~339$-$4, XTE~J1550$-$564,
 and XTE~J1859$+$226) while in the intermediate and 
steep power-law states, we are able to present a general pattern of behavior for the origin of 
radio emission in these two states that could be important for understanding the accretion-ejection
coupling very close to the black hole event horizon.   

\end{abstract}


\section{Introduction}

\xt\ is a soft X-ray transient discovered on 2001 September 5 (MJD 52157, Remillard 2001) by 
the All-Sky Monitor on-board the {\em Rossi X-ray Timing Explorer} ({\em RXTE}/ASM). On the
next day, pointed {\em RXTE}/PCA (Proportional Counter Array) observations confirmed the 
ASM detection of \xt\, with an X-ray 
spectrum typical of a black hole candidate in the hard state (Markwardt, Swank, \& Smith 2001). 
This was confirmed by further analysis of the power density spectra (Revnivtsev \& Sunyaev 2001;
Wijnands, Miller, \& Lewin 2001).  During the course of the outburst, \xt\ went into all the
canonical X-ray spectral states (Rossi et al. 2003) that are typical of the population of black 
hole candidates (BHCs) (Belloni 2003; McClintock \& Remillard 2004). High frequency quasi-periodic 
oscillations (QPOs) have been reported by Homan et al. (2003). The possible detection of a broad 
iron K$\alpha$ emission (e.g. Miller et al. 2002) may suggest 
that \xt\ is a maximal Kerr black hole. In addition, short ($\sim$ 100 s) X-ray flares and long 
time-scale oscillations have been reported by Tomsick et al. (2003b) during the decay to 
quiescence.

The optical counterpart was discovered by Castro-Tirado et al. (2001) with the 0.6-m
optical telescope at Lake Tekapo, New Zealand.  The identification was later confirmed by 
Groot et al. (2001) and Augusteijn, Coe, \& Groot (2001). The radio counterpart was discovered 
with the Australia Telescope Compact Array (ATCA) located in Narrabri, Australia 
(Groot et al. 2001).  Further optical observations
of \xt\ in quiescence with the Magellan 6.5-m and VLT 8-m telescopes revealed the orbital
parameters of this system (Orosz et al. 2004). The companion star is of spectral type 
K3V to K5V, and the orbital period of the system is 0.3205 days. The mass  function 
of 2.73 $\pm$ 0.56 \sm\ and the lower limit on the inclination angle of 50 $\pm$ 3\degr\
(in contradiction to what was originally proposed by Sanchez-Fernandez et al. 2002) 
give an upper limit of 7.3 \sm\ for the mass of the compact object.
The primary in the \xt\ system is likely a black hole, and in fact, the existing data 
suggests that it could be a black hole with a mass of only 4 \sm\ (Orosz et al. 2004).

In this paper, we report on X-ray and radio observations of \xt\ spread over the
entire outburst. In \S2, we describe the observations and the outburst evolution.
We then discuss the properties of radio emission from \xt\ in each of the states 
covered by our observations, including the hard state, the steep power-law state 
(we also discuss the nature of the intermediate state), the thermal dominant state 
and the state transition. Our conclusions are summarized in \S4.  Our X-ray state 
definitions are nearly the same as the new definitions described in the review by 
McClintock \& Remillard (2004). In this work, we follow McClintock \& Remillard 
(2004) by using the names hard state (hereafter HS) and thermal dominant (TD) state rather 
than the previous names (low-hard state and high-soft state, respectively) to 
designate the two best known X-ray states.  The difference between our definitions 
and those of McClintock \& Remillard (2004) concerns the intermediate and very high 
states  which could possibly be various instances of the same state (e.g. Homan 
et al. 2001, Belloni 2003).  McClintock \& Remillard (2004) divided these two flavors 
of the intermediate/very high state as follows:  The soft (c.f. photon index $>$ 2.4) 
flavor, which has a steep power-law spectrum and was therefore called the the steep 
power-law (SPL) state.  In the SPL state, low and high frequency QPOs are usually
detected.  The interpretation of the intermediate state in McClintock \& Remillard 
(2004) is more ambiguous as it is defined in terms of combinations of the other 
spectral states.  In this work, we define the intermediate state (IS) as having properties 
between the HS and SPL states so that the intermediate state is essentially
a hard flavor of the intermediate/very high state mentioned previously.

\section{Observations and Outburst Overview}

\subsection{Radio observations}

During the X-ray outburst of \xt, we conducted eight continuum radio observations with the Australia 
Telescope Compact Array (ATCA) located in Narrabri, New South Wales, Australia. The ATCA synthesis 
telescope is an east-west array consisting of six 22 m antennas. The ATCA uses orthogonal polarized 
feeds and records full Stokes parameters.  We carried out observations mostly at 4800 MHz (6.3 cm) 
and 8640 MHz (3.5 cm) with the exception of the first two observations, for which we also made 
measurements at 1384 MHz (21.7 cm) and 2496 MHz (12.0 cm). For the first observation, due to the 
large uncertainty in the X-ray position at that time, the radio counterpart of \xt\ was outside the 
primary beam of the telescope at 4800 and 8640 MHz. We performed observations in various array 
configurations: 6B (baselines ranging from 214 m to 5969 m), 6D (77 m to 5878 m), 0.750D (31 m to 
4469 m) and EW352 (31 m to 4438 m), in order of decreasing spatial resolution.  An additional 
observation (6A configuration) was conducted on 2003 December 21 while \xt\ was in quiescence. 
We did not detect \xt\ with a three sigma upper limit of 0.3 mJy at 4800 and 8640 MHz.

The amplitude and band-pass calibrator was PKS~1934$-$638, and the antenna's gain and phase 
calibration, as well as the polarization leakage, were derived from regular observations of the 
nearby (less than a degree away) calibrator PMN~1646$-$50.  The editing, calibration, Fourier 
transformation, deconvolution, and image analysis were performed using the MIRIAD software package 
(Sault \& Killen 1998).  An observing log as well as the ATCA flux densities of \xt\ can be found 
in Table~1. The dates of our ATCA observations are indicated in Figs.~1 and 2 in order to 
illustrate how they are related to the X-ray state of \xt.

\subsection{X-ray observations}

In order to have an long-term view of the X-ray behavior of \xt, we used the publically 
available X-ray data from the {\em RXTE}/ASM (Levine et al. 1996). The 1.5-12 keV ASM 
light-curve is plotted in Figure~1. In addition, we also used Proportional Counter Array
(PCA) and High Energy X-ray Timing Experiment (HEXTE) data from our pointed observations
as well as observations available from the {\em RXTE} archive.  The procedure for reduction 
of these data can be found in Tomsick et al. (2003b; 2004).  We used these data to extract 
count rates in the 3-200 keV band as a function of time.  We define a hard color as the 
ratio of the HEXTE count rate (20-200 keV) over the 3-20 keV PCA count rate.  We then 
constructed a hardness-intensity diagram (HID), similarly to Homan et al. (2003) and 
Rossi et al. (2003).  We conducted a more detailed analysis of the X-ray energy spectrum 
for the {\em RXTE} observations closest in time to the eight radio observations.  Using 
the XSPEC (version 11) software for spectral analysis, we fitted the PCA+HEXTE spectra, 
primarily to determine the fluxes in several energy bands (see Table 1 and Figures 5 and 6).
In all cases, the spectral continuum is well-described by a disk-blackbody (Makishima 
et al. 1986) plus power-law or cutoff power-law model, and this is typical of BHC systems.  
We also accounted for interstellar absorption, and we fixed the column density to the
value of $N_{\rm H} = 6\times 10^{21}$ cm$^{-2}$ measured by the {\em Chandra X-ray
Observatory} (Tomsick et al. 2004).  For some observations, this continuum model left 
significant residuals near the iron K$\alpha$ complex (6-10 keV), but we obtained 
acceptable fits with reduced-$\chi^2 < 1.0$ after including an iron emission line and a smeared iron 
edge (Ebisawa et al. 1994).  The final model we used to determine the X-ray fluxes is 
also described in detail in, e.g. Tomsick, Corbel, \& Kaaret (2001).  

\subsection{The 2001-2002 X-ray outburst of \xt\ }     

To fully understand the radio properties of \xt\ during its outburst, we first need to 
characterize the X-ray states of \xt\ as a function of time.  For that purpose, Fig.~1
shows the {\em RXTE}/ASM 1.5--12 keV and {\em RXTE}/PCA+HEXTE 3--200 keV light-curves, 
as well as the evolution of the {\em RXTE}/ASM (3--12 keV/1.5--3 keV) hardness ratio; 
the arrows indicate the dates of the ATCA radio observations. In addition, the HID in 
Fig.~2 adds complementary information on the outburst evolution; the diamonds highlight 
the radio observations. \xt\ moved counter-clockwise in the HID during the whole outburst.
As can be seen from these two figures, the scheduling of the radio observations provides
a sampling of very different X-ray states of \xt.  In Figure 1, we indicate the time
intervals used by Homan et al. (2003) and Rossi et al. (2003) with Roman numerals. 

After its discovery on 2001 September 5 (MJD 52157) by {\em RXTE}/ASM (Remillard 2001), 
the first {\em RXTE} pointed observation occurred on 2001 September 6. The {\em RXTE} 
observations (up to September 9) are consistent with a BHC in the HS  with a 
strong band limited noise component in the power density spectra and energy spectra 
dominated by a power-law component of photon index $\sim$ 1.6 with exponential cut-off 
(Revnivtsev \& Sunayev 2001; Wijnands et al. 2001).  The rest of the bright outburst 
phase has been described by Rossi et al. (2003 and private communication) and Homan et 
al. (2003), and we outline their conclusions below (for the decay phase, see Kalemci 
et al. 2003 and Tomsick et al. 2003b, 2004). Starting around September 9, a gradual 
softening of the spectrum occurs up to October 5 (MJD 52187).  From Sept. 9 
to Sept. 20 (MJD 52161-52172), \xt\ is characterized by a smooth softening of its energy 
spectrum (with an evolution of the power-law photon index from 1.5 to 2.2) and the 
frequency of the QPO increased from $\sim$ 1 to 9 Hz. During the period from  Sept. 20 to Oct. 5 (MJD 
52172-52187), the photon index of the power-law component saturates to 
a value of $\sim$ 2.2. The rms variability and the frequency of the QPO become more 
erratic with oscillations around their maxima. The high frequency QPOs are only observed 
during this portion of the outburst (Homan et al. 2003).  The accretion disk and 
power-law flux components give similar contributions to the total flux, which makes 
this interval very typical of the SPL state. We note that the photon index 
does not exactly fulfill the criteria (not greater than 2.4) of McClintock \& Remillard 
(2004) for a SPL state, but as it the part of the outburst with the steepest power-law, 
we will consider it as a SPL state for the rest of the paper. The properties of \xt, 
between Sept. 9 to Sept. 20, would then be consistent with an IS as defined above.  
From Oct. 5 to Nov 19 (MJD 52187-52232), the contribution from the accretion 
disk dominates the energy spectra and the level of rms variability is very low as is 
typical of the TD state.  \xt\ returned back to the HS after Nov 19 (MJD 52232) 
as illustrated by the hardening of the spectrum and the increased rms 
variability (Kalemci et al. 2003; Rossi et al. 2003). 
 
To summarize, after a brief (but we can not exclude that the outburst was ongoing for many 
days before the discovery of the source) initial HS, \xt\ underwent a smooth 
transition to the IS, followed by transitions to the  SPL state, the TD state and then back to the HS.  Our 
radio observations occurred as followed:  \# 1 and 2 during the initial HS (however,
 \# 2 is very close from the transition to the IS), \# 3 and 4 during the SPL state
(however, \# 4 is very close from the transition to the TD state), \# 5 and 6 during the TD state
and \# 7 and 8 during the final HS. A ninth observation was performed later when \xt\ 
was back in quiescence. We now describe the properties of \xt\ along these X-ray states.

\section{Radio emission from \xt: Results and Discussion}     

We now focus on the properties of the radio emission.  We obtained radio coverage
during the initial and final hard states, the steep power-law state and the thermal 
dominant state.  By looking at Table 1 and Fig. 3, where the radio light curve of \xt\ 
is plotted, we note that radio emission from \xt\ is detected during the first radio 
observation on September 7 with a spectrum consistent with being flat between 1384 
and 2496 MHz. The day after, the radio flux density increased by almost a factor of two
with similar spectral characteristics.  On September 24, we observed the source at a 
much fainter radio flux ($\sim$ 0.8 mJy), and the source disappeared below the sensitivity 
level on October 5 with three sigma upper limits of 0.21 and 0.18 mJy at 4800 and 8640 MHz,
respectively. Compared to the brightest level of radio emission observed on September 8, 
this indicates a significant quenching (of more than a factor 25) of the radio emission. 
Surprisingly, radio emission is again observed (Obs. \# 5 and 6) at the mJy level in the 
TD state (contrary to expectations based on observations of other BHCs, 
e.g., Fender et al. 1999; Corbel et al. 2000).  During the final two radio observations, 
the behavior of the source may be consistent with the behavior during the first two 
observations.  For the rest of the paper, we define the radio spectral index, $\alpha$, 
as S$_\nu$ $\propto$ $\nu^\alpha$, where S$_\nu$ is the radio flux density and $\nu$ is
the frequency.

The best position of the radio counterpart to \xt\ (with the radio source fitted as a point-like 
source) is: $\alpha$(J2000) = 16$^h$50$^m$00.96$^s$ and $\delta$(J2000) = --49\degr 57\arcmin 
44.60\arcsec\ with an absolute positional uncertainty of 0.25\arcsec, mostly due to the 
uncertainty on the phase calibrator position. All radio observations (when the radio source 
is detected) are consistent with a location of the radio counterpart at this position. This 
constitutes the most accurate position for \xt, and is in agreement with the one derived from 
optical observations (Castro-Tirado et al. 2001) and {\em Chandra} observations (Tomsick 
et al. 2004).

\subsection{The Initial and Final Hard states}

\subsubsection{Radio emission from a compact jet}

Radio observations \# 1, 2, 7 and 8 were performed while \xt\ was in the HS. Both 
the initial and final hard X-ray states are therefore covered. Once again, the overall 
properties of the hard state radio emission are broadly consistent with those that have
been observed in other BHCs in a similar X-ray state (Corbel et al. 2000; Fender 2001): 
A level of radio emission of a few mJys with a radio spectrum that is almost flat. Such 
characteristics are believed to originate from a self-absorbed conical outflow or compact 
jet (e.g. Blandford \& K$\ddot{\mathrm{o}}$nigl 1979 or Hjellming \& Johnson 1988), 
similar to the one directly resolved from \cx\ by Stirling et al. (2001).  We do not 
detect linear polarization from the compact jet of \xt, with our best 3 $\sigma$ upper 
limit of 4.0\% or 4.7\% at 4800 or 8640 MHz, respectively.  Such limits are consistent 
with previous detections at lower levels (e.g. Corbel et al. 2000 for \gx).

For the purpose of our discussion, we have calculated the radio spectral indeces, and
these are included in Table 1.  Despite being consistent with flat ($\alpha$ $\sim$ 0), 
the radio spectrum (e.g. Fig. 4) seems less inverted at high radio frequencies
than is typical for BHC systems. 
Later, we will come back to the last 
observation on December 4, which shows a very unusual spectrum. It is interesting to 
note that the radio spectrum on September 8 (Observation \# 2: Figure 4) shows a 
turnover at lower frequencies.  This could be due to free-free absorption by a thermal 
plasma, and, indeed, a fit to the spectrum with a power law and free-free absorption 
(S$_\nu$ = S$_0$ $\nu^\alpha$ exp(-$\tau$ $\nu^{-2.1}$), where S$_0$ is the amplitude 
at 1 GHz, $\alpha$ is the spectral index of the un-absorbed spectrum and $\tau$ is the 
free-free optical depth at 1 GHz) describes the data sufficiently well with S$_0$ = 
8.59 $\pm$ 0.07 mJy, $\alpha$  = --0.29 $\pm$ 0.04 and $\tau$ = 1.32 $\pm$ 0.02. The 
opacity is in the range of values obtained by Fender (2001) for the 1989 outburst 
of V404~Cyg (GS~2023$+$338) (Han \& Hjellming 1992). We note that similarly, during 
the first detection of XTE~J1859$+$226 in 1999, the radio spectrum also shows absorption 
at low frequency (Brocksopp et al. 2002) while in the HS.  If the radio emission 
arises from a compact jet (as is usually observed in the HS), then it is unlikely 
that synchrotron self-absorption is responsible for the observed absorption at low-frequency
as this emission originates from large scale regions (see Fender 2001). The nature of 
the putative thermal absorbing plasma is unclear (higher ISM density, remnant of past 
activity, etc.).  Alternatively (but probably less plausibly), the radio spectrum could 
be caused by two components: A flat component from the compact jet as well as a second
component from an optically-thick ejection event.  This may be a possibility as radio 
observation \# 2 occurred very close to the hard to intermediate state transition.  
However, this is not favored by the fact that the radio spectra are all consistent with 
having the same intrinsic spectral index, even during the final HS (with the 
exception of observations \# 8).  Also, as discussed below, if an ejection event occurred 
for \xt, it probably took place during the intermediate to steep power-law state transition 
(section 3.2).  

We note that the radio spectrum in observation \# 8 is very steep ($\alpha$ $\le$ --1.3), 
and this is unusual for a black hole in the HS. It is not clear if this is related 
to the jet/ISM interaction mentioned below (likely not, as the spectrum in observation 
\# 7 looks similar to the initial HS) or possibly to the X-ray oscillation behavior 
observed ten days later (Tomsick et al. 2003b). 

As discussed above, we obtained the best constraint on the spectral index during 
the September 8 observation with $\alpha$  = --0.29 $\pm$ 0.04. The reason the radio 
spectrum may be less inverted than usual is still unclear. This may be related to the inclination angle of the 
jet as usually lower inclination angles lead to flatter spectra (e.g. Falcke 1996). For 
example, the radio to millimeter spectrum of the compact jet of \cx\ (a low inclination 
system) is almost flat, i.e. $\alpha$ $\sim$ 0 (Fender et al. 2000). However, even with 
a low inclination angle of the jet, it is almost impossible to obtain such a negative
spectral index. Furthermore, optical observations indicate that the inclination angle of 
the orbital plane in \xt\ is at least 50\degr\ (Orosz et al. 2004), so this explanation 
does not work for \xt\ unless the compact jets are strongly misaligned with the orbital 
plane (as might be the case in few systems, Maccarone 2002). Additionally, we note that even 
in a system such as 4U~1543$-$47 with a low inclination angle (20.7 $\pm$ 1.0\degr, 
Orosz et al. in prep.), the radio spectrum of the compact jet (e.g. $\alpha$ = 0.08 $\pm$ 0.04) 
is still slightly inverted (Kalemci et al. 2004b). We also note that the spectral index 
also vary within a single source (e.g. \gx, Corbel et al. 2000),  so it is unlikely that the
inclination angle is responsible for this less inverted radio spectrum. 

According to  Hjellming \& Johnston (1988), a less inverted radio spectrum would also be 
expected in the case of slowed lateral expansion by an external medium (i.e. a compact jet 
with a narrower opening angle). In addition, due to the longitudinal pressure gradient, 
the bulk Lorentz factor will increase along the jet axis. If the observer is looking into 
the jet boosting cone, then one would expect to see an increase in low frequencies radio 
emission (which originates far from the base of the jet) and therefore possibly a much 
less inverted radio spectrum (e.g. Falcke 1996). Obscuration of part of the receding jet
may also contribute to the nature of the spectrum. In any case, a combination of all these 
effects may be in place in \xt. It is also clear that future studies of the evolution of 
the radio spectral index of the compact jets in BHC systems are important as they 
may shed light of the geometry of the system.

If the measured spectral index is correct, then it is likely that the contribution from 
the compact jet at higher frequencies will not be significant. As illustrated in \gx\ 
and \xte, the spectrum of the compact jet extends to shorter wavelengths, with a transition 
to optically thin regime in near infrared (Corbel et al. 2001; Corbel \& Fender 2002). In 
that case, it is unlikely that an infrared re-flare would have been detected during the 
soft to hard state transition as in \xte, \gx\ or 4U~1543$-$47 (Jain et al. 2001; Buxton 
\& Bailyn 2003).  The contribution in X-rays, as direct synchrotron emission (c.f. Markoff, 
Falcke, Fender 2001; Markoff et al. 2003) may also be negligible (see section 3.1.2), however
a contribution as synchrotron self-Compton (SSC) emission from the base of the compact jet 
(e.g. Markoff  \& Nowak 2004) can not be ruled out at this stage.

\subsubsection{On the radio/X-ray correlation}

While in the HS, black hole candidates display a strong correlation between 
their radio and X-ray emission. This was first observed in \gx\ (Corbel et al. 2000; 
2003) over more than three orders of magnitude in X-ray flux and almost down to its 
quiescence level. Gallo, Fender, \& Pooley (2003) found a similar correlation for 
\gs\ (V404 Cyg), but, interestingly, they show that all BHCs in the the  HS
behave similarly, i.e. their data are consistent with a universal relation between 
radio and X-ray luminosities.  They also observed a relatively small scatter of 
approximately one order of magnitude in radio power.  In fact, the scatter could
be even smaller if we account for the recent distance estimate, in the range 
between 6 and 15 kpc, for \gx\ by Hynes et al. (2004) as this will bring \gx\ 
closer to \gs, and the small scatter could imply low bulk Lorentz factors 
($<$ 2) for the compact jets.  Interestingly, this correlation also seems to 
hold for a large sample of super-massive black holes if one take into account 
the mass of the black hole as an additional correction (Merloni, Heinz, \& 
di Matteo 2003; Falcke, K\"ording, \& Markoff 2004). 

In order to see if \xt\ fits into this picture, we looked at the relationship
between the X-ray and radio flux levels for \xt. Despite our efforts to get 
quasi-simultaneous X-ray and ATCA observations (see Table 1), this was not 
possible for the final HS as \xt\ was in the solar exclusion zone for 
{\em RXTE}. In Figure 5, we plot the X-ray flux in various bands during a portion 
of the decay of the 2001 outburst (see also Tomsick et al 2003b, 2004).  After 
the transition to the HS (on MJD 52232, Kalemci et al. 2004a), the decay 
is smooth enough that interpolation of the X-ray flux at the time of the radio 
observation is sufficient to obtain an estimate.  We note that an increase of 
the decay rate is observed at the end of the outburst (Figure 5) with a spectrum 
that gets harder with time (a smooth variation of the photon index: 1.91 $\pm$ 
0.02 on MJD 52235.6 to 1.35 $\pm$ 0.23 on MJD 52267.9), very similar to the decay 
of XTE~J1908$+$094 during its 2003 outburst (Jonker et al. 2004).

In Figure 6, we have plotted the radio flux density at 4.8 GHz versus the 
unabsorbed 2--11 keV X-ray flux (in Crab units) scaled to a distance of 1 kpc 
(similar to Gallo et al. 2003), assuming a distance of 3 kpc for \xt. 
This distance gives luminosity estimates during the state transition 
consistent with other BHCs (Maccarone 2003).  In Fig. 6, we have also plotted the 
best-fit function obtained by Gallo et al. (2003) using their datasets, i.e. 
S$_{\mathrm{radio}} = \mathrm{k} \times (S_x)^{+0.7}$ with k = 223 $\pm$ 156 mJy. Although 
we have a sample of only four data points, we can clearly see that they all lie 
significantly below (by a factor 20) the best-fit line of Gallo et al. (2003).  
The determination of the slope of the power-law function linking the radio and 
X-ray emission is very uncertain due to our limited sample of data points 
(as well as the unusual radio spectrum in observation \# 8 and the fact that two observations 
took place very close to state transitions). The reason why the normalization 
in \xt\ is significantly lower (the source is less radio loud or more X-ray loud)
than in other BHCs (Gallo et al. 2003) is still unclear. It may be related to 
the fact that direct synchrotron X-ray emission (and possibly even SSC) from 
the compact jets is likely not dominant in the case of \xt\ for a reason which 
still remains to be explained. The lower normalization in \xt\ could be related 
to the fact that the X-ray regime may be dominated by thermal Comptonization of disk
photons in the corona.
We note that the distance to \xt\ may be 
larger than 3 kpc, but this will only make \xt\ more anomalous
relative to the other BHCs.

\subsection{Radio emission in the intermediate and steep power-law states}       

\subsubsection{The steep power-law state of \xt }

As outlined above, after a few days in the hard state, \xt\ entered
the intermediate state  and then went into the steep power-law state. 
In the IS, the X-ray spectrum softens gradually until the 
photon index reaches a value of $\sim$2.2 (Rossi et al. 2003), whereas the QPO 
frequency increases from 1 Hz to 8.5 Hz. Then (SPL state), the photon index, 
and the QPO frequency (it is not clear if this is the same type of QPO) 
oscillate around their saturation value.  This behavior is more pronounced for 
the QPO frequency as it fluctuates along with the broad-band variability. The 
slow variations of the photon index could be related to cooling of the corona by 
an increase in the number of soft disk photons (the inner accretion disk could 
get closer to the black hole).  Significant variations are also observed in the 
total disk flux contribution (Rossi et al. 2003) and may indicate that the 
accretion disk has reached the innermost stable circular orbit (ISCO) and is 
oscillating around this value. This interval (SPL state) also corresponds to the period over 
which high frequency QPOs are detected and therefore confirms that the disk is 
very close to the black hole (Homan et al. 2003).

We conducted two radio observations during the SPL state
of \xt\ in 2001. The first one resulted in the  detection of \xt\ at a faint level 
of $\sim$0.8 mJy (the spectral index is not well constrained: $\alpha$ = --0.13 
$\pm$ 0.86).  The second observation did not result in a detection of \xt\ with 
3 $\sigma$ upper limits of 0.18 mJy at 8640 MHz and 0.21 mJy at 4800 MHz, indicating 
a significant (more than a factor 25) quenching of radio emission compared to the 
initial HS.

\subsubsection{The steep power-law state of \xte\ and XTE~J1859$+$226}

Very few soft X-ray transients have been observed at radio frequencies during the 
intermediate or steep power-law states. Moreover, when this state has been observed, 
the radio emission has usually been dominated by the decaying optically thin 
synchrotron emission arising from jet ejections that occurred at or near state 
transitions prior to the source entering the intermediate or steep power-law state.
Thus, when the emission from the jet ejections is detected, it is decoupled from 
the black hole system, implying that the observed radio emission is not an intrinsic 
property of the intermediate or steep power-law states as the emitting electrons 
are already far from the system. 

There is only one case for which the radio observations sampled the intrinsic 
properties of the intermediate or steep power-law states: \xte\ during its 
reactivation in 2000 (Corbel et al. 2001).  Interestingly, the properties of 
the radio emission from \xte\ were quite similar to what we observe now in 
\xt: Indeed, for \xte, the first detection showed an optically thin spectrum 
with a well-constrained spectral index of $\alpha$ = --0.45 $\pm$ 0.05, while, 
later, the radio emission was quenched. The optically thin component was 
interpreted as synchrotron emission arising from relativistic plasma during the 
state transition.  Such small ejection events are frequently observed during 
state transitions (e.g. see Fender et al. (1999) or Corbel et al. (2000) for 
the 1998 outburst of \gx\ or Brocksopp et al. (2004) for the 2003 outburst of 
XTE~J1720$-$318). We note that the spectacular massive ejection events (with 
bright radio emission and a radio core that is usually resolved on a time-scale 
of weeks; e.g. Mirabel \& Rodr\' \i guez (1994) for \grs) may be related to 
sharper state transition possibly related to a huge increase in the accretion 
rate in the inner part of the accretion disk or maybe to a different black hole 
parameter such as the spin. 

Furthermore, we highlight the fact that during the 1999 outburst of XTE~J1859$+$226, 
Brocksopp et al. (2002) reported the detection of flaring radio emission from this 
black hole in a soft state, which is unexpected in the canonical high/soft (or TD) 
state (e.g. Fender et al. 1999). In that case, the radio flaring emission was 
clearly associated with spectral hardening of the X-ray spectrum.  However, their 
definition of a soft state is not clear. Indeed, the hard X-ray light-curve in 
Brocksopp et al. (2002) revealed a very significant level of hard X-ray emission 
up to at least MJD 51490, which is very uncommon for a TD state (e.g. McClintock 
\& Remillard 2004). In addition, Cui et al. (2000) reported the detection of high 
frequency QPOs around MJD 51468, which is a characteristic of the SPL state 
(McClintock \& Remillard 2004), this is also favored by the large fraction of the 
X-ray flux in the power-law component (Hynes et al. 2002), which is also quite steep. 
Kalemci (2002), analyzing 
the {\em RXTE}/PCA data of XTE~J1859$+$226 after MJD 51515, describes the spectral 
state evolution as: A TD state from MJD 51515 up to MJD 51524, after which the 
system was found in the IS for the remaining PCA observations. Based on the above, 
it seems likely that the TD state in XTE~J1859$+$226 did not start before MJD 51490. 
This is confirmed by Markwardt (2001) as this work indicates that the disc component 
became dominant only after MJD 51487. Therefore, the X-ray state in which Brocksopp 
et al. (2002) detected flaring radio emission from XTE~J1859$+$226 was most likely 
the SPL state (or less possibly an IS) and not a TD state.  Therefore, the observed 
radio flaring behavior in XTE~J1859$+$226 must be related to the behavior of BHCs 
while in the SPL state. This conclusion has also been drawn by
Fender, Belloni \& Gallo (2004).

\subsubsection{\xt\ and the origin of radio emission in the steep power-law state}

With our new observations, \xt\ is therefore the third source for which the intrinsic 
radio properties of the SPL state have been sampled, and the observed 
behavior is consistent with what has been found in \xte\ (Corbel et al. 2001). The 
detection of the faint radio component in the SPL state on September 24 (observation 
\# 3) may be related to an ejection event at the time of the transition from the hard 
to intermediate states (on September 9).  However, as there is significant time 
(15 days) between the transition and the detection of the faint radio component, 
we consider two alternative explanations.

First, an interesting comparison can be made with the 2002-2003 outburst of \gx\ as 
a strong radio flare has been observed with ATCA (Gallo et al. 2004). In Fig. 7, 
we have plotted the ASM light-curve and hardness ratio (similar to Fig. 1 for \xt) 
for the initial part of the outburst.  As \gx\ and \xt\ have similar hydrogen column 
density (e.g. Miller et al. 2004), we can directly compare their hardness ratios, 
which are indeed very similar (Figs. 1 and 7). After an initial HS, \gx\ 
made a transition to an IS around MJD 52400. Then, the spectrum softened up to 
approximately MJD 52409, when \gx\ is found in the SPL state, the evolution of
the states became more complicated later on in the outburst (T. Belloni, private 
communication). The spectral state evolution (during the beginning of the outburst) 
in \gx\ is therefore identical to \xt. The radio flare observed by Gallo et al. (2004)
started on 2002, May 14 around 13:00 (MJD 52409.042), reaching its maximum 6 hours 
later.  The arrow in Figure 7 indicates that the radio flare started once the 
softening of the X-ray spectrum ended, i.e. it would correspond to the transition 
from the IS to the SPL state. Based on the above, and comparing their hardness 
ratios, we can say that if a radio flare (and hence massive plasma ejection) 
occurred in \xt, then this happens at the transition between IS and SPL state. 
In that case, the observed radio emission on September 24 
would be the end of the decay of this flare. This may constitute the ejected 
materials that may be interacting later with the interstellar medium (section 3.3) 
as in \gx\ (Gallo et al. 2004).
 
Alternatively, an equally plausible explanation is the following. The transition 
from the IS to the SPL states seems to correspond to a period over which 
the inner radius of the accretion disk reaches the ISCO, and this would be valid 
for most of the SPL state as outlined above in section 3.2.1.  As the X-ray properties 
of the SPL state favor an ``unstable'' accretion disk (Rossi et al. 2003), they may 
suggest that the optically thin synchrotron component observed on September 24 
could be related to an ejection of a very small portion of the inner accretion 
disk or corona. In that case, it would be very similar to the radio flaring 
behavior observed in XTE~J1859$+$226 during its SPL state in 1999 or to the 
behavior observed in \grs\ (but at much slower rate). The comparison of \grs\ 
with BHCs in canonical X-ray states is not straightforward, but the X-ray states 
A, B (soft) and C (hard) may be similar to intermediate and very high states
(Reig, Belloni \& van der Klis 2003). Oscillations between states A, B and C are 
correlated with radio flaring activity (e.g. Klein-Wolt et al. 2002) with 
stronger radio emission in the spectrally hard state C, whereas the soft states 
are never associated with bright radio emission. 

These two possible explanations for the origin of radio emission in the IS 
and SPL state could be combined as follows.  For \gx\ and probably for \xt,
the transition from IS to SPL state is associated with an ejection event (with 
a radio spectrum characteristic of optically thin synchrotron emission) that 
decays on a time-scale of hours. After the transition, the accretion disk settles 
down close to the ISCO. At that time, as the spectrum hardens, accretion disk or
coronal material can be ejected from the system (as in XTE~J1859$+$226), resulting 
in weak radio flares. If the radio observation takes place between two flares, 
then no radio emission would be observed. Indeed, this was also the case for 
\gx: After the major radio flare associated with the IS-SPL state transition, 
several radio flares were observed during the SPL state (Gallo et al. 2004), 
and, interestingly, for at least one SPL state radio observation (on 2002 June 9 
= MJD 52434), the radio emission was quenched (by a factor $>$ 45 compared to 
the initial HS).

Indeed, we can now try to see if the 2000 radio observations of \xte\ fit into 
this framework. For that purpose, we have plotted in Figure 8 the ASM hardness 
ratio and light-curve for the 2000 outburst of \xte.  Similarly, after an initial 
HS (e.g. Rodriguez, Corbel, \& Tomsick 2003; Rodriguez et al. 2004),
the source went into the SPL state with gradual softening of the X-ray spectrum
in between (very similar to the IS of \xt\ and \gx\ mentioned above). The radio 
flare was observed again after the softening was over (Corbel et al. 2001). 
Again, the high frequency QPOs (Miller et al. 2001) were only reported in the 
SPL state (between MJD 51663 and MJD 51672), i.e after the end of the softening. The 
second radio observation of \xte\ occurred later in the SPL state and showed 
that the radio emission was quenched so that either the observation occurred
between two flares or perhaps there was no flare at all. The X-ray spectral 
evolution  (Rodriguez, Corbel, \& Tomsick 2003) may even suggest that the ejected
material could be originating from the corona. The outburst evolution 
of XTE~J1859$+$226 (Figure 9) could again be included in this picture: After an 
initial HS and a softening, a massive radio flare occurred; then, as 
the source became spectrally harder, several weaker flares took place.

\subsubsection{On the nature of radio emission in the intermediate state}

At this stage, it is not clear if the general pattern presented in the previous 
subsection only concerns the SPL state. Perhaps the IS, with slower 
evolution and a significant level of hard X-ray emission (like the IS in \xt), 
is not associated with radio emission. In fact, it is interesting to note that before the major radio 
flare observed in 2002 (see Figure~2 in Gallo et al. 2004), a stable level of 
radio emission with a flux of $\sim$ 12 mJy and a flat spectrum (between 4.8 
and 8.6 GHz) was observed in \gx.  So, as \gx\ was in the IS at that time, 
this would indicate that the compact jet could possibly survive during the 
IS but could be destroyed very quickly, i.e. on time-scale of hours. 
Similarly, a non-zero level of radio emission with a flat spectrum is also observed 
in XTE~J1859$+$226 prior to its major radio flare in 1999 (Brocksopp et al. 2002). 
Again this detection would be consistent with the presence of the compact jet in the
IS of XTE~J1859$+$226. 
If this interpretation is correct (i.e. the compact jet exists in the IS), this would 
be an important clue for understanding 
the inflow-outflow coupling close to the event horizon of a black hole. For 
that purpose, it would be very important to monitor the radio properties 
of BHCs during the IS or SPL state in order to constrain the geometry of 
these BHC systems during the various X-ray states. In any
case, this suggestion seems to be confirmed by Fender et al. (2004), who studied
similar datasets (including also \grs, but not \xt) and drawn similar conclusions
regarding the nature of radio emission in the IS and SPL state.

\subsubsection{Quenched radio emission at few percent Eddington luminosity}

As a final remark, in Figure 6, we observe that the radio observations in the 
initial HS and in the SPL state occurred at the same unabsorbed X-ray 
flux, which corresponds to a level of 4 Crab ($\sim$ 8\% of the Eddington 
luminosity for a 4 \sm\ black hole at 3 kpc), (if we used an upper limit of 
7 \sm, then it would correspond to about 4.6\% of the Eddington luminosity). 
This picture is qualitively consistent with the behavior of \gx, \cx\ and \gs\
(Gallo et al. 2003), indicating that the quenching of the compact jet occurs 
at an almost fixed fraction (few percent) of the Eddington luminosity so that
there is a correlation between the mass of the black hole and the X-ray flux
where the transition occurs.  We note that if the mass of a black hole is 
known, then a measurement of the X-ray flux for which quenching occurs
could constitute an independent distance estimate (or vice versa). 

\subsection{Surprising Detection of radio Emission in a Thermal Dominant state}

The last radio observations that we discuss in this paper are those (\# 5 and 6) 
that were conducted during the TD state. Radio 
emission is observed (Figures 3 and 10) at a level of $\sim$ 1 mJy with a spectrum 
that is consistent with optically thin synchrotron emission (but the spectral 
index is not well constrained).  These detections are contrary to what would 
have been expected in the TD state, which has always (in previous observations
of BHCs) been associated with quenched radio emission: e.g. \gx\ (Fender et al. 
1999; Corbel et al. 2000) and \cx\ (Gallo et al. 2003; Tigelaar et al. 2004). 
Therefore, these detections do not fit with the standard view, and they constitute 
a certain surprise.  As discussed in section 3.2, the radio flaring emission 
observed in a soft state of XTE~J1859$+$226 (Brocksopp et al. 2002) has to be 
related to the behavior of BHC while in the SPL state.  Regarding the case of 
\xt, the HID (Fig. 2) indicates that the detection of radio emission in the 
TD state in not related to spectral hardening at all, as the X-ray emission 
stays very soft during this period with no hard component. So, an alternative 
explanation must be found, and we will concentrate on two possibilities. 

First, we consider the possibility that the radio emission is still related to 
the compact jet.  As demonstrated in the case of \gx\ (Fender et al. 1999; 
Corbel et al. 2000), the TD state is associated with a quenching of the compact
jet by at least a factor of 25.  According to Meier, Koide, \& Uchida (2001), 
the quenching would be the result of a weaker poloidal magnetic field in 
geometrically thin accretion disk. Migliari et al. (2004) reported the detection 
of radio emission in two atoll-type neutron star X-ray binaries while they were 
in a soft (banana) X-ray state. They suggested that this could be related to 
an interaction of the magnetic field of the neutron star with the accretion 
disk. However, such explanation does not work in the case of \xt, which is 
likely a black hole based on its X-ray properties and also its mass function 
(Orosz et al. 2004).  Maybe the mass of the black hole in \xt\ (that may be 
smaller than typical stellar mass black hole) is an important parameter that 
could set the level of quenching when the spectrum gets soft.  However, 
as the compact jet was quenched by a factor of $>$ 25 in the SPL state 
(observation \# 4), similar to other BHCs (e.g. \gx\ Fender et al. 1999),
it seems likely that the observed radio emission in the TD state of \xt\
does not originate from the compact jet. 

The second possibility that we consider now is that the observed radio emission 
is the result of the interaction of material previously ejected from the system 
with the interstellar medium, similarly to what has been observed for \xte\ 
(Corbel et al. 2002; Tomsick et al. 2003; Kaaret et al. 2003) and for \gx\
(Gallo et al. 2004). The observed radio spectrum would be consistent with this 
interpretation (optically thin synchrotron emission, but it should be kept in
mind that the spectral index is not well constrained). In addition, we also 
observed that within an observation, the flux is varying (decaying) on time-scale 
of hours. For example, in observation \# 5, the radio flux density drops (during 
the observation) from 1.74 $\pm$ 0.11 mJy to 0.85 $\pm$ 0.11 mJy at 4800 MHz 
and from 1.09 $\pm$ 0.11 mJy to 0.80 $\pm$ 0.15 mJy at 8640 MHz.  There also seems
to be some variations for the observation \# 6.
This is contrary to what has been found previously in \xte\ with the slow decay 
of radio emission (on a time-scale of a week) due to the jet/ISM interaction 
(Corbel et al. 2002; Corbel et al. in prep.). However, the variations 
of radio emission in the large scale jet of \gx\ was much faster than in the case 
of \xte.  The origin of the ejected material could be related to the flaring 
behavior in the SPL state or more likely during the IS/SPL state transition (as 
discussed in section 3.2), similarly to \gx.  If this interpretation is correct, 
the contribution to the X-ray spectrum of the interaction of the jet with the ISM 
(as in in \xte) would not be detectable, as the X-ray spectrum would likely be 
dominated by the thermal emission from the accretion disk. In any case, this may 
suggest that the reactivation of particle acceleration during collisions with the 
interstellar medium may be a common occurrence in microquasars.

\section{Conclusions}

\xt\ was discovered in September 2001 and then underwent transitions between various
X-ray spectral states while we observed the source at radio frequencies. We can 
summarize our conclusions as follows. In the hard state, the radio emission of 
\xt\ can be interpreted (like other BHCs) as arising from a self-absorbed compact 
jet. However, there seems to be some indication that the radio spectrum is less 
inverted than in other sources. In addition, \xt\ seems to be more X-ray loud when 
compared to other black hole candidates observed at similar radio flux density. 
This could possibly indicate that \xt\ is dominated in the X-ray regime by Comptonization of the disk photons 
in  the corona with negligible (if any) contribution from the compact jet at high energies 
(X-ray, optical, and possibly even in the infrared).  With the observations performed 
in the steep power-law state and using the existing data from other BHCs, we 
conclude that the transition from IS to SPL state is likely associated with a (more 
or less) massive ejection event that decays on a time-scale of hours.  In addition, 
weaker radio flares (and hence ejection events) may be observed in the SPL state
associated with X-ray spectral hardening.  If a radio observation took place between 
two flares or if no flare occurred at all, then no radio emission would be detected. 
For the IS itself, the detection of radio emission (with a flat spectrum) prior to 
the major flare of XTE~J1859$+$226 in 1999 and \gx\ in 2002 (associated to an IS/SPL state transition) may 
suggest that the compact jet can survive in the IS, and perhaps this is due to
the fact that the flux of soft X-rays is lower in the IS than in the SPL state.
In the TD state, we have surprisingly detected a significant amount of varying 
radio emission that we interpret as the interaction of previously ejected materials 
with the neighboring environment (ISM or remnant of past activity), similar to what 
has been observed in \xte.  In that case, such events may be more common than 
previously thought. Our conclusions regarding the nature of radio emission 
along the various spectral states of \xt\ and a possible extension to
black hole candidates in general are summarized in Table 2).
All of this points to the fact that it is extremely important 
to intensively monitor the radio properties of BHCs along the various X-ray states 
in order to shed light on the accretion - ejection coupling close to the black hole
event horizon. 

\acknowledgements

The Australia Telescope is funded by the Commonwealth of Australia for operation as a national
Facility managed by CSIRO. {\em RXTE}/ASM results are provided by XTE/ASM team at MIT. We thanks Bob
Sault, Dave McConnell and the ATCA TAC for allowing these observations at the right time
and that have sample various X-ray states.
SC acknowledge useful and interesting discussions with Heino Falcke, Elena Gallo, Sera Markoff, Mike Nowak, and Jerry Orosz.
SC would like to thank Dick Hunstead and Duncan Campbell-Wilson for
providing informations on the MOST observations that help scheduling the ATCA observations
and Bryan Gaensler, Jim Lovell for conducting some of the radio observations.
We also warmly thank Sabrina Rossi and Tomaso Belloni for providing informations on their PCA data
analysis of \xt\ and \gx, that helps defining the state evolution.
JAT acknowledges partial support from NASA grant NAG5-13055.

\newpage

\begin{table*}[ht]
\caption{\label{tab_log} Observing log and results }
\begin{minipage}{\linewidth}
\hspace{-1cm}
\tiny
\begin{tabular}{ccccccccc}
\hline \hline
        &   \multicolumn {8}{c}{Radio Observations\footnote{Upper limits are given at the 3 sigma confidence level.}} \\
\hline
	  & 1  & 2     & 3     & 4     & 5     & 6     & 7     & 8     \\
\noalign{\smallskip}
\cline{2-9}
\noalign{\smallskip}
Date (MJD)\footnote{Radio observation midpoint} & 52159.94   & 52160.81   & 52177.01   & 52187.67   & 52195.85   & 52204.25
  & 52241.52   & 52247.63   \\
Calendar   & 2001:09:07 & 2001:09:08 & 2001:09:24 & 2001:10:05 & 2001:10:13 & 2001:10:21 & 2001:11:27 & 2001:12:04 \\
Time on source (hr)\footnote{The first number is for the observation at 1384 and 2496 MHz (if noted), otherwise it is related to the observations at 4800  and 8640 MHz} & 0.66 ; 1.09  & 1.16 ; 0.99  & 1.29 & 4.10 & 3.75 & 1.80 & 6.56 & 2.73  \\
Array configuration & 6B & 6B  & 0.750D & EW352 & EW352 & EW352 & 6D & 6D \\
\noalign{\smallskip}
\cline{2-9}
\noalign{\smallskip}
        &   \multicolumn {8}{c}{Flux density (mJy)} \\
\noalign{\smallskip}
\cline{2-9}
\noalign{\smallskip}
 1384 MHz & 2.7 $\pm$ 0.3 & 4.08 $\pm$ 0.20  &    ....  &    ....  &    ....  &    ....  &    ....  \\
 2496 MHz & 2.3 $\pm$ 0.2 & 5.30 $\pm$ 0.15  &    ....  &    ....  &    ....  &    ....  &    ....   \\
 4800 MHz &  .... & 5.28 $\pm$ 0.10 & 0.83 $\pm$ 0.10 & $<$ 0.21      & 1.29 $\pm$ 0.07 & 0.78 $\pm$ 0.10 & 1.21 $\pm$ 0.06
&  0.75 $\pm$ 0.10 \\
8640 MHz  &  .... & 4.48 $\pm$ 0.10 & 0.77 $\pm$ 0.10 & $<$ 0.18        & 0.91 $\pm$ 0.10 & 0.34 $\pm$ 0.09 &1.15 $\pm$ 0.06
 &$<$ 0.30         \\
Spectral index & -0.27 $\pm$ 0.31 & -0.29 $\pm$ 0.04\footnote{Spectral index obtained if we fit the radio spectrum with a 
power-law and thermal free-free absorption. If we used the two higher frequencies, a spectral index of -0.28 $\pm$ 0.14 is 
deduced.} & -0.13 $\pm$ 0.86 & .... & -0.59 $\pm$ 0.56 & -1.41 $\pm$ 1.36 & -0.09 $\pm$ 0.18 & $<$ --1.3 \\
\noalign{\smallskip}
\hline \hline
\noalign{\smallskip}
        &   \multicolumn {8}{c}{X-ray Observations} \\
\noalign{\smallskip}
\hline
\noalign{\smallskip}
X-ray state & HS & HS & SPL & SPL & TD & TD & HS & HS \\
\noalign{\smallskip}
\cline{2-9}
\noalign{\smallskip}
        &   \multicolumn {8}{c}{Unabsorbed 2-11 keV flux (in unit of 10$^{-9} \rm \, erg \,s^{-1} \,cm^{-2}$))} \\
\noalign{\smallskip}
\cline{2-9}
\noalign{\smallskip}
Flux & 10.40 $\pm$ 0.10  &  10.96 $\pm$ 0.11 & 10.47 $\pm$ 0.10 & 10.31 $\pm$ 0.10 & 10.04 $\pm$ 0.10 & 7.25 $\pm$ 0.07 & 1.75 $\pm$ 0.08 & 1.13 $\pm$ 0.06 \\
\noalign{\smallskip}
\hline
\end{tabular}
\end{minipage}
\end{table*}

\begin{table*}[hb!]
\vspace{-0.5cm}
\caption{\small Summary of the properties of radio emission along the various X-ray states 
in \xt\ and in black hole binary systems in general (see also Fender et al. 2004 
for the general case). We note that in any state observed after the IS to SPL 
state transition, radio emission from the interaction of the massive ejection 
event with the interstellar medium may contribute to the observed level of radio emission (if unresolved). }
\begin{center}
\begin{tabular}{lcc}
\noalign{\smallskip}
\tableline \tableline
\noalign{\smallskip}
X-ray state 			&  \multicolumn {2}{c}{Origin of the radio emission}	\\	
\noalign{\smallskip}
\cline{2-3}
\noalign{\smallskip}
				&	in \xt	&    in black hole candidates \\
\noalign{\smallskip}
\tableline
\noalign{\smallskip}
Hard state  			& Self-absorbed compact jet 		& Self-absorbed compact jet \\
\noalign{\smallskip}
Intermediate state 		& No radio observation			& Self-absorbed compact jet \\
\noalign{\smallskip}
IS to SPL state transition      & No radio observation 			& Massive ejection event    \\
\noalign{\smallskip}
Steep power-law state		& Decay of massive ejection event  	& Small ejection events  and/or \\
				& or small ejection event + quenching	&  quenched radio emission \\
\noalign{\smallskip}
Thermal dominant state		&	Interaction jet/ISM ?		& Quenched radio emission  \\
\noalign{\smallskip}
\tableline \tableline
\end{tabular}
\end{center}
\end{table*}

\begin{figure}
\plotone{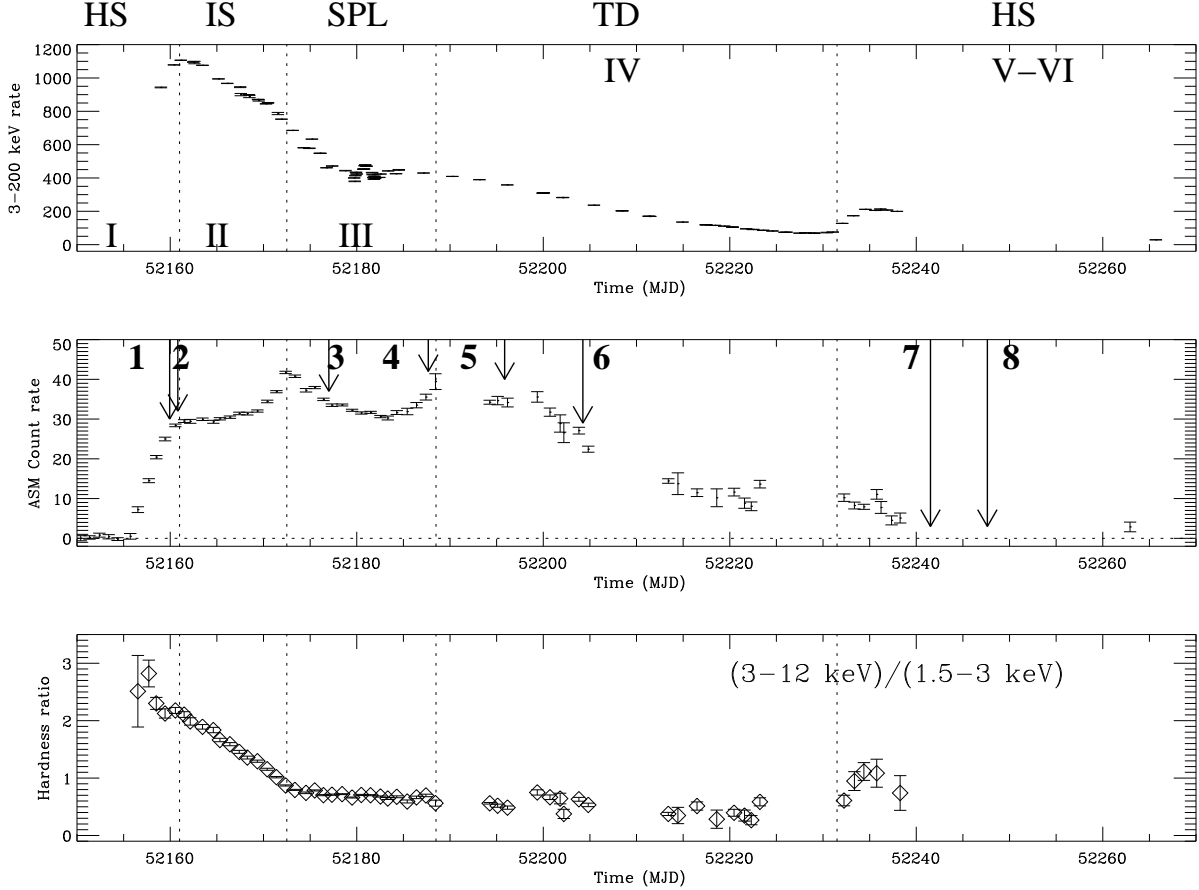}
\caption{\it Top: {\em RXTE} (PCA + HEXTE) 3--200 keV count rate light-curve (daily averaged) of \xt\ during its 
2001-2002 outburst. Middle: 1.5--12 keV {\em RXTE}/ASM count rate light-curve. Bottom: Evolution 
of the ASM hardness ratio (3--12 keV/1.5--3 keV) during the whole outburst. The vertical dotted 
lines indicate the transition between the various X-ray states: HS (hard state), IS (intermediate 
state), SPL (steep power-law state) and TD (thermal dominant state). The roman numerals in the 
top panel illustrate the various time intervals used by Homan et al. (2003) and Rossi et al. 
(2003) for their X-ray analysis. The arrows (with a number) indicate when our radio observations 
have been performed. We note that the high frequency QPOs have only been detected in interval III 
(Homan et al. 2003).
}
\end{figure}

\begin{figure}
\plotone{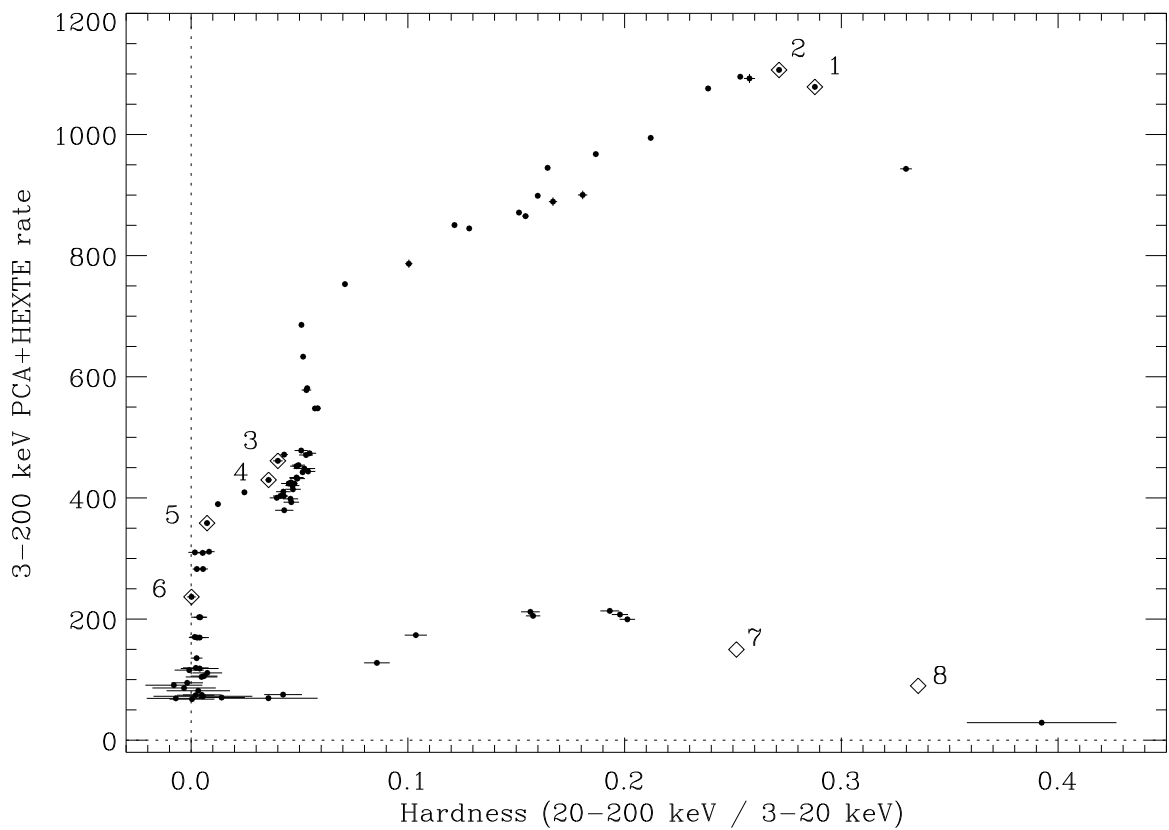}
\caption{\it  Hardness Intensity Diagram (HID) for the outburst of \xt\ similar to the one used by 
Homan et al. (2003).  The diamonds (with associated number) indicate the period of simultaneous radio 
and X-ray observations. For observations \# 7 and 8, the position is only indicative as \xt\ could 
not be observed by {\em RXTE} due to its proximity to the Sun. However (see text and Figure 5), 
their positions are likely to be approximately correct.}
\end{figure}

\begin{figure}
\plotone{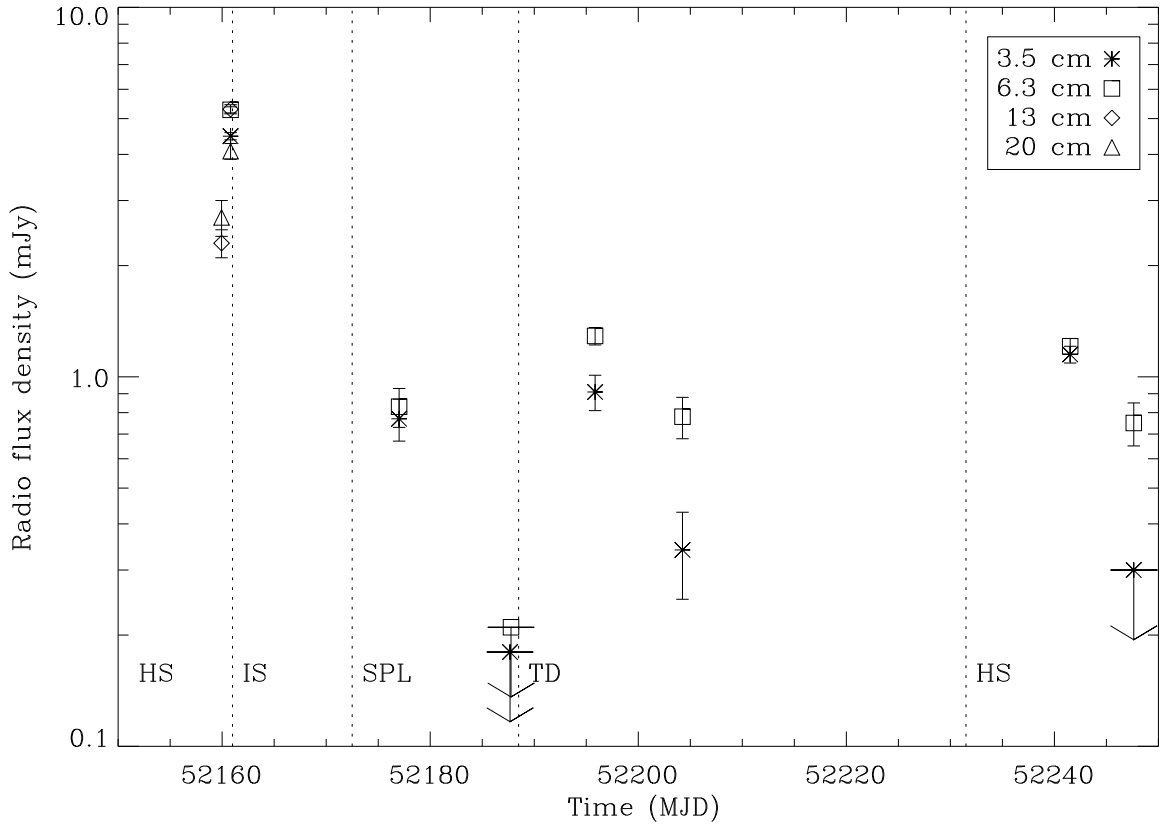}
\caption{\it  Radio light-curve of \xt\ during its outburst in 2001. The vertical lines define
the state transitions (see also Figure 1). Upper limits are plotted at the three sigma 
confidence level.}
\end{figure}

\begin{figure}
\plotone{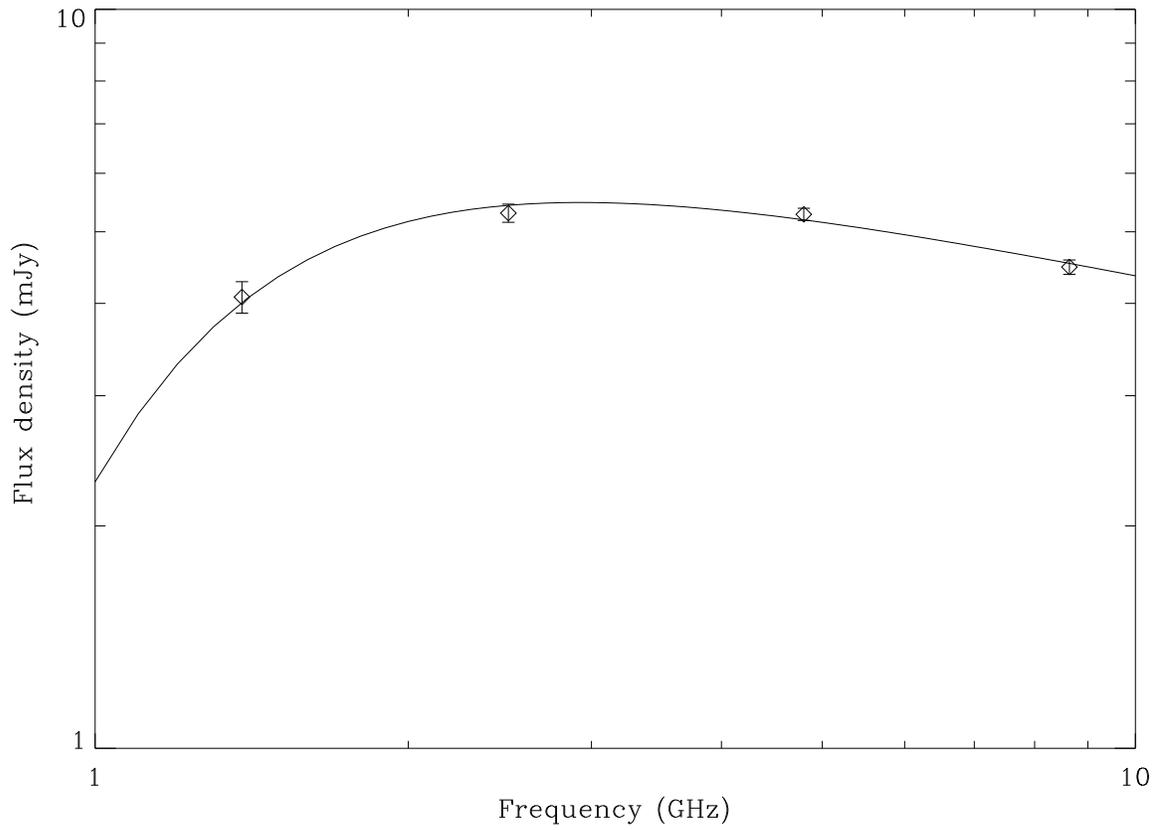}
\caption{\it  Radio spectrum of \xt\ for the observation on 2001 September 8 (\# 2). The 
continuous line is the fit to spectrum with a power-law and thermal free-free absorption. }

\end{figure}

\begin{figure}
\plotone{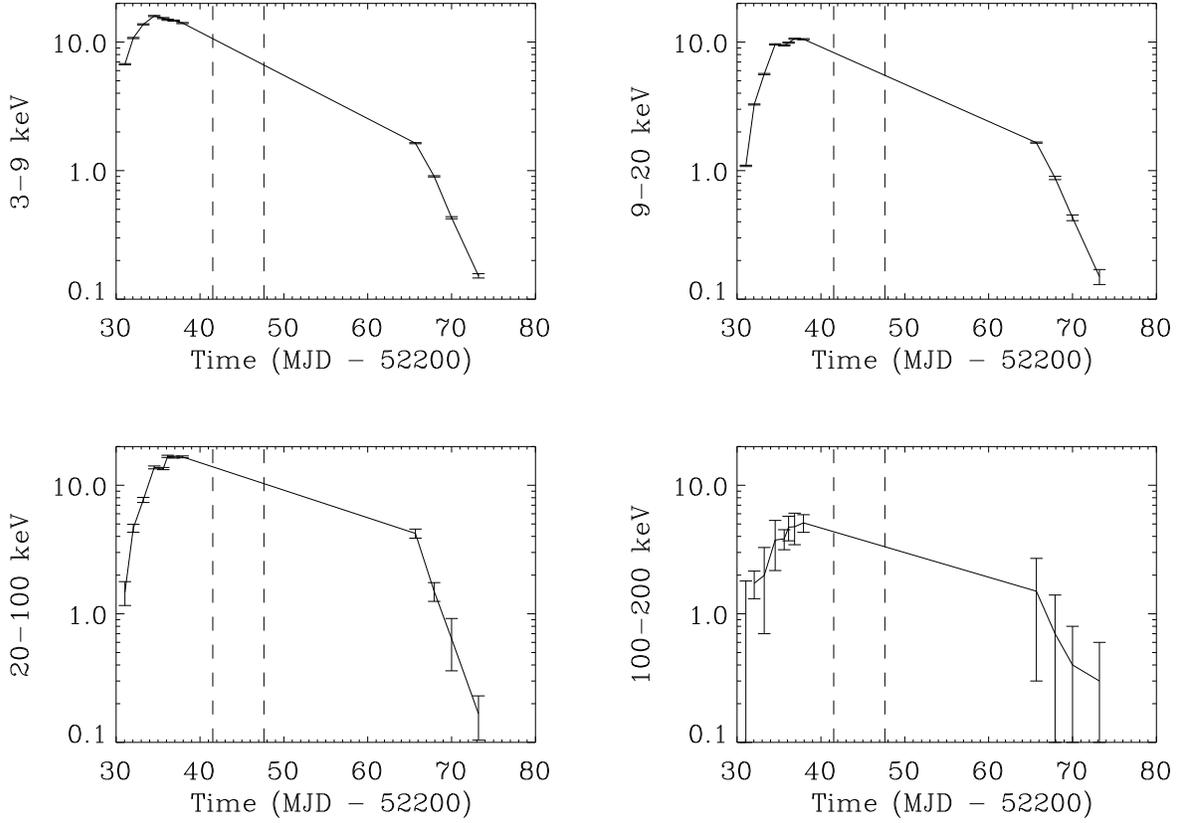}
\caption{\it  Evolution of the X-ray flux (in units of 10$^{-9}$ erg s$^{-1}$ cm$^{-2}$) in 
various energy bands during the decay phase. The transition from TD state to HS occurred on 
MJD 52232. The dashed vertical lines indicate when radio 
observations \# 7 and 8 were performed. These figures give a feeling for the precision
of our X-ray flux estimates (for radio observations \# 7 and 8) based on an interpolation 
of the decay trend.}
\end{figure}

\begin{figure}
\plotone{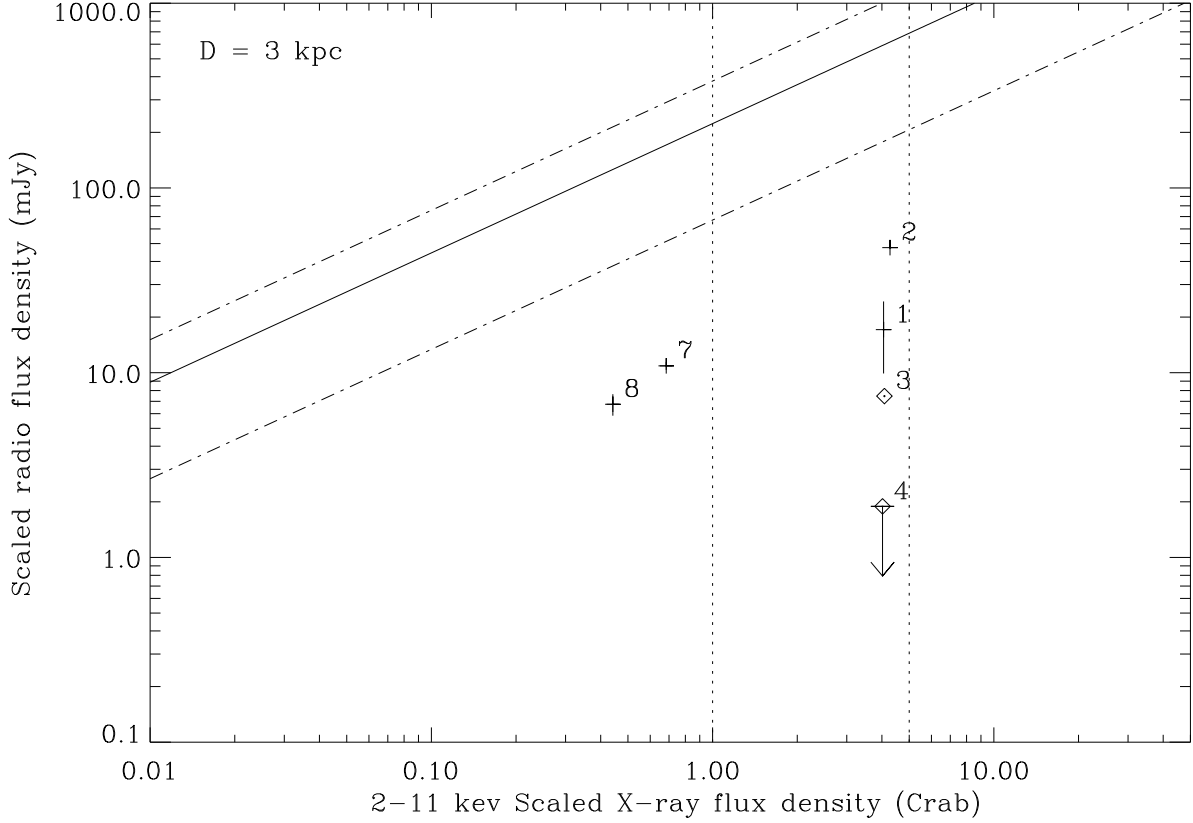}
\caption{\it  Radio flux density (mJy) at 4.8 GHz of \xt\ during the HS and SPL state versus the unabsorbed 
2-11 keV X-ray flux (in Crab units) scaled to a distance of 1 kpc (this assumes that \xt\ is 
located at 3 kpc from the Earth). As described in the text, the X-ray flux for the observations 
\# 7 and 8 are from an interpolation of the decay trend.  The best fit function (continuous line, 
with its associated error (dot-dashed lines)) obtained by Gallo et al. (2003) for \gx\ and \gs\ 
is also plotted. The vertical dotted lines indicate the level of 2\% and 10\% Eddington luminosity 
for a 4 \sm\ black hole. } 
\end{figure}

\begin{figure}
\plotone{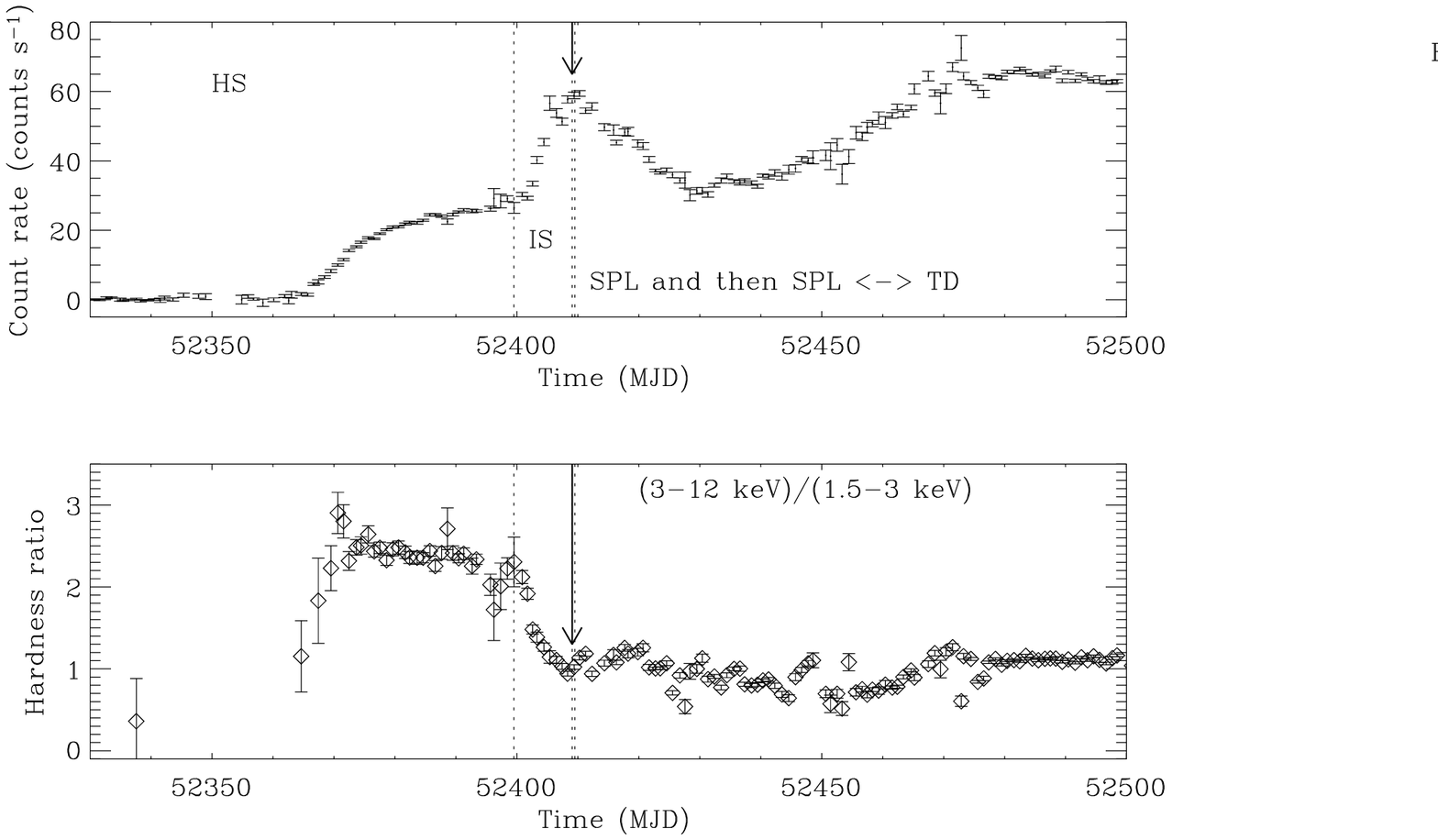}
\caption{\it Evolution of the {\em RXTE}/ASM count rate and hardness ratio (similar to Figure 1) 
for the initial part of the 2002 outburst of \gx.  Again, the X-ray spectral states sampled
(T. Belloni, private communication) are indicated.  The arrow marks the time of the major radio 
flare observed by Gallo et al. (2004) in May 2002.}
\end{figure}

\begin{figure}
\plotone{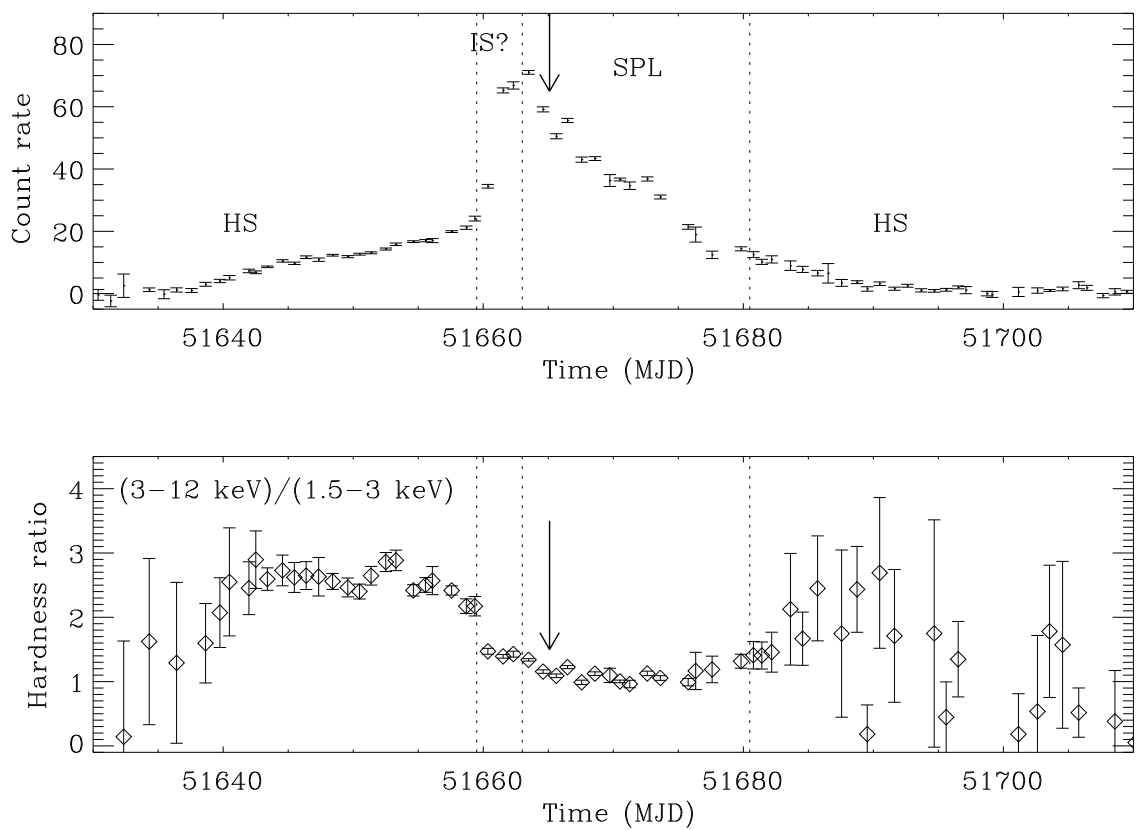}
\caption{\it  Same as Figure 7 but for the 2000 outburst of \xte. The arrow indicates the date 
of the radio observations performed by Corbel et al. (2001) during which they probably detected 
the end of the radio flare associated to the state transition. The high frequency QPOs are only 
detected during the SPL state. }
\end{figure}

\begin{figure}
\plotone{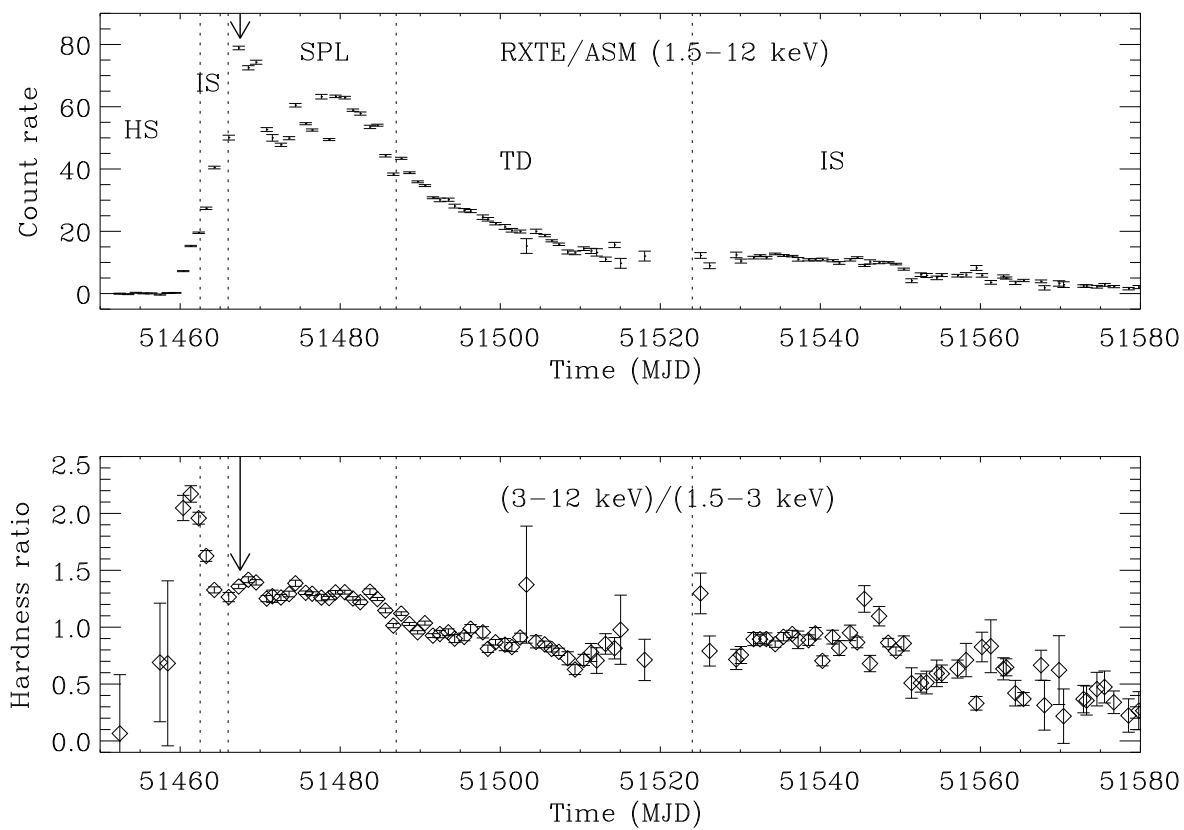}
\caption{\it  Same as Figure 7 but for the 1999 outburst of XTE~J1859$+$226. The arrow indicates 
the date of the major radio flare detected by Brocksopp et al. 2002). The two first vertical lines for the HS
and IS are approximately indicative of the period of state transitions.}
\end{figure}

\begin{figure}
\plotone{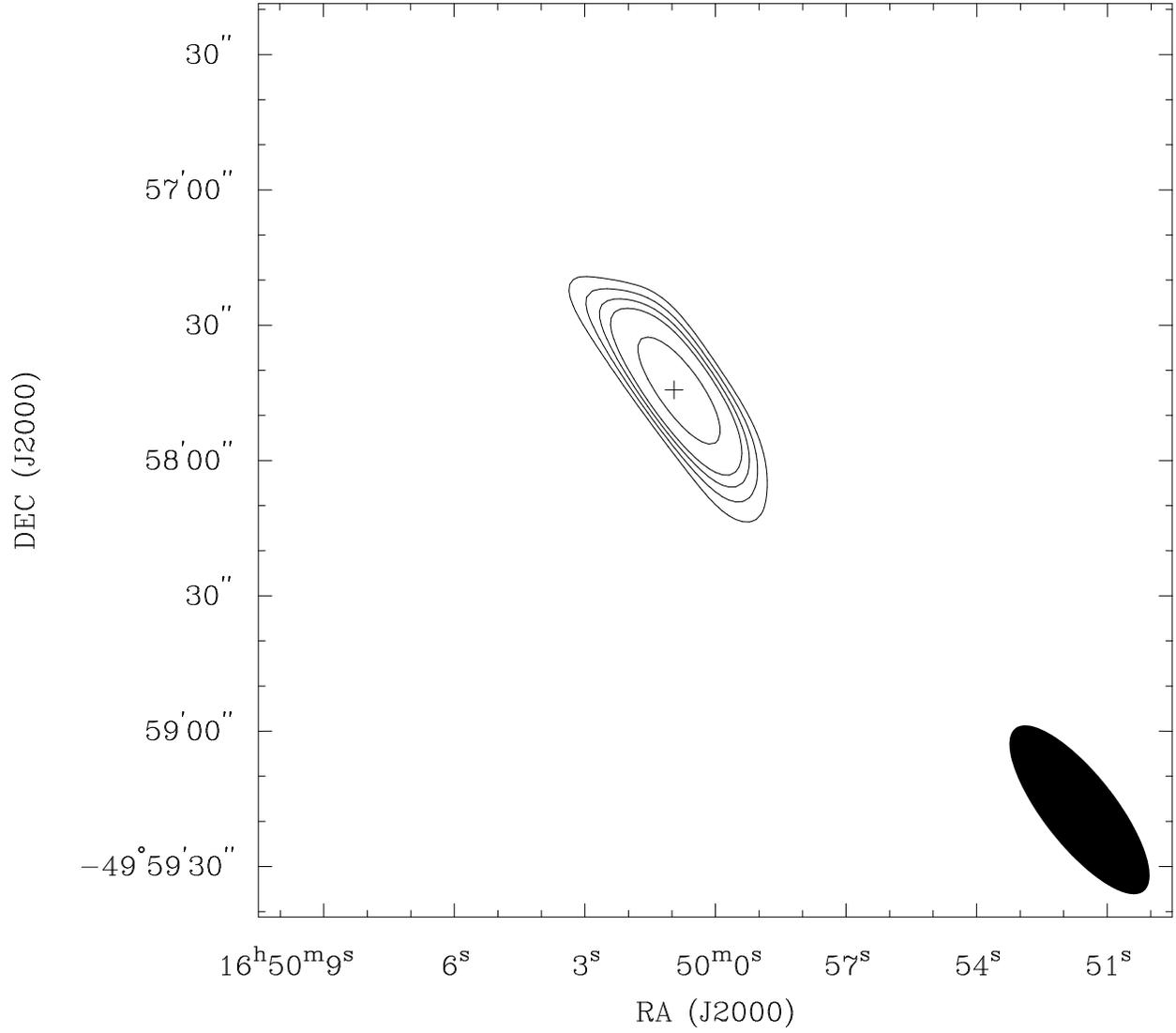}
\caption{\it  Radio emission at 8640 MHz during an observation performed on 
MJD 52195 (observation \# 5) in a TD state. Contours are at 3, 4, 5 , 7 and 9 times
the r.m.s. level of 0.10 mJy/beam.}
\end{figure}

\clearpage


\newpage

\begin{thebibliography}{99}   

\bibitem[Augusteijn, Coe, \& Groot 2001]{aug01} Augusteijn, T., Coe, M., \& Groot, P.\ 2001, \iaucirc, 7710


\bibitem[Belloni  2003]{bel03}Belloni, T. 2003, Proc. "The Restless High-Energy Universe", astro-ph/0309129            


\bibitem[Blandford \& K$\ddot{\mathrm{o}}$nigl 1979]{bla79} Blandford,
R. D., \& K$\ddot{\mathrm{o}}$nigl, A. 1979, \apj, 232, 34

\bibitem[Brocksopp et al. 2002]{bro02} Brocksopp, C., et al.\ 2002, \mnras, 331, 765 

\bibitem[Brocksopp et al. 2004]{bro04} Brocksopp, C., Corbel., S., Fender, R. P., Rupen, M., Sault, R., Tingay, S.J., 
	Hannikainen, D., O'Brien, K.  2004, \mnras, to be submitted

\bibitem[Buxton \& Bailyn 2003]{bux03}Buxton, M., \& Bailyn, C. 2003, Proc. ``X-Ray Timing 2003: Rossi and Beyond'', astro-ph/0312392

\bibitem[Castro-Tirado et al. 2001]{cas01} Castro-Tirado, A.~J., Kilmartin, P., Gilmore, A., Petterson, O., Bond, I., Yock, P., \& 
Sanchez-Fernandez, C.\ 2001, \iaucirc, 7707

\bibitem[Corbel et al. 2000]{cor00} Corbel, S., Fender, R.\ P.,
Tzioumis, A.\ K., et al. \ 2000, \aap, 359, 251

\bibitem[Corbel et al. 2001]{cor01} Corbel, S. et al. 2001, \apj, 554, 43                   

\bibitem[Corbel \& Fender 2002]{cf02} Corbel, S.~\& Fender, R.~P.\ 2002, \apjl, 573, L35 

\bibitem[Corbel et al. 2002]{cor02} Corbel, S., Fender, R.~P., Tzioumis, A.~K., Tomsick, J.~A., Orosz, J.~A., Miller, J.~M., 
Wijnands, R., \& Kaaret, P.\ 2002, Science, 298, 196 

\bibitem[Corbel et al. 2003]{cor03} Corbel, S., Nowak, M.~A., Fender, R.~P., Tzioumis, A.~K., \& Markoff, S.\ 2003, \aap, 400, 1007 

\bibitem[Cui et al. 2000]{cui00} Cui, W., Shrader, C.~R., Haswell, C.~A., \& Hynes, R.~I.\ 2000, \apj, 535, L123 


\bibitem[Ebisawa et al. 1994]{ebi94} Ebisawa, K., et al.\ 1994, \pasj, 46, 375 


\bibitem[Falcke 1996]{fal96} Falcke, H.\ 1996, \apjl, 464, L67 

\bibitem[Falcke, K\"ording, \& Markoff 2004]{fal04}Falcke H., K\"ording E., Markoff S., 2004, A\&A, 414, 895

\bibitem[Fender et al. 1999]{fen99} Fender R.\ P. et al.\  1999, \apj, 519, L165                      

\bibitem[Fender et al. 2000]{fen00} Fender, R.~P., Pooley, G.~G., Durouchoux, P., Tilanus, R.~P.~J., \& Brocksopp, C.\ 2000, \mnras, 
312, 853 

\bibitem[Fender 2001]{fen01} Fender R.\ P. 2001, \mnras, 322, 31                      

\bibitem[Fender, Belloni, Gallo 2004]{fen04} Fender R.\ P., Belloni, T., Gallo, E. 2004, submitted

\bibitem[Groot et al. 2001]{gro01} Groot, P., Tingay, S., Udalski, A., Miller, J. 2001, \iaucirc, 7708

\bibitem[Gallo, Fender, \& Pooley 2003]{gal03} Gallo, E., Fender, R.~P., \& Pooley, G.~G.\ 2003, \mnras, 344, 60 

\bibitem[Gallo et al. 2004]{gal04} Gallo, E., Corbel, S., Fender, R.~P., Maccarone, T.~J., \& Tzioumis, A.~K.\ 2004, \mnras, 347, L52 

\bibitem[Hjellming \& Johnston 1988]{hje88} Hjellming R. M., \& Johnston K. J. 1988, \apj, 328, 600                

\bibitem[Homan et al. 2001]{hom01} Homan, J., Wijnands, R., van der Klis, M., Belloni, T., van Paradijs, J., Klein-Wolt, M., Fender, 
R., \& M{\' e}ndez, M.\ 2001, \apjs, 132, 377 

\bibitem[Homan et al. 2003]{hom03} Homan, J., Klein-Wolt, M., Rossi, S., Miller, J.~M., Wijnands, R., Belloni, T., van der Klis, M., 
\& Lewin, W.~H.~G.\ 2003, \apj, 586, 1262 

\bibitem[Hynes et al. 2002]{hyn02} Hynes, R.~I., Haswell, C.~A., Chaty, S., Shrader, C.~R., \& Cui, W.\ 2002, \mnras, 331, 169 

\bibitem[Hynes et al. 2004]{hyn04} Hynes, R.~I., Steeghs, D., Casares, J., Charles, P.~A., \& O'Brien, K.\ 2004, \apj, 609, 317

\bibitem[Jain et al. 2001]{jai01} Jain, R.~K., Bailyn, C.~D., Orosz, J.~A., McClintock, J.~E., \& Remillard, R.~A.\ 2001, \apjl, 554, 
	L181 

\bibitem[Jonker et al. 2004]{jon04} Jonker, P.G., Gallo, E., Dhawan, V., Rupen, M., Fender, R.P., Dubus, G. 2004, \mnras, 351, 1359

\bibitem[Kaaret et al. 2003]{kaa03} Kaaret, P., Corbel, S., Tomsick, J.~A., Fender, R., Miller, J.~M., Orosz, J.~A., Tzioumis, A.~K., 
\& Wijnands, R.\ 2003, \apj, 582, 945 

\bibitem[Kalemci 2002]{kal02} Kalemci, E. 2002, {\em Ph.D. Thesis}, University of California, San Diego

\bibitem[Kalemci et al. 2003]{kal03} Kalemci, E., Tomsick, J.~A., Rothschild, R.~E., Pottschmidt, K., Corbel, S., Wijnands, R., 
Miller, J.~M., \& Kaaret, P.\ 2003, \apj, 586, 419 

\bibitem[Kalemci et al. 2004a]{kal04a} Kalemci, E., Tomsick, J.~A., Rothschild, R.~E., Pottschmidt, K., \& Kaaret, P.\ 2004a, \apj, 603, 231 

\bibitem[Kalemci et al. 2004b]{kal04b} Kalemci, E., Tomsick, J.~A., Buxton, M., Bailyn, C., Rothschild, R.~E., Pottschmidt, K., Corbel, S., Brocksopp, C., Kaaret, P.\ 2004b, \apj, to be submitted

\bibitem[Klein-Wolt et al. 2002]{kle02} Klein-Wolt, M., Fender, R.~P., Pooley, G.~G., Belloni, T., Migliari, S., Morgan, E.~H., \& 
van der Klis, M.\ 2002, \mnras, 331, 745 

\bibitem[Levine et al. 1996]{lev96} Levine, A.~M., Bradt, H., Cui, W., Jernigan, J.~G., Morgan, E.~H., Remillard, R., Shirey, R.~E., 
\& Smith, D.~A.\ 1996, \apjl, 469, L33 

\bibitem[Maccarone 2002]{mac02} Maccarone, T.~J.\ 2002, \mnras, 336, 1371 

\bibitem[Maccarone 2003]{mac03} Maccarone, T.~J.\ 2003, \aap, 409, 697 

\bibitem[Makishima et al. 1986]{mak86} Makishima, K., Maejima, Y., Mitsuda, K., Bradt, H.~V., Remillard, R.~A., Tuohy, I.~R., 
Hoshi, R., \& Nakagawa, M.\ 1986, \apj, 308, 635 

\bibitem[Markoff, Falcke, \& Fender 2001]{mar01} Markoff, S., Falcke,
H., \& Fender, R.\ 2001, \aap, 372, L25
 
\bibitem[Markoff et al.  2003]{mar03} Markoff, S., Nowak, M., Corbel, S., Fender, R., \& Falcke, H.\ 2003, \aap, 397, 645

\bibitem[Markoff \& Nowak 2004]{mar04} Markoff, S., \& Nowak, M. 2004, \apj, in press, astro-ph/0403468

\bibitem[Markwardt 2001]{markw01a} Markwardt, C.\ 2001, Astrophysics and Space Science Supplement, 276, 209 

\bibitem[Markwardt, Swank, \& Smith 2001]{markw01} Markwardt, C., Swank, J., \& Smith, E.\ 2001, \iaucirc, 7707

\bibitem[McClintock \& Remillard 2004]{mcc04} McClintock, J.~E., \& Remillard, R.~A. 2004, in Compact Stellar X-ray sources, eds.
	W.~H.~G. Lewin \& M. van der Klis, (Cambridge: Cambridge University Press), in press, astro-ph/0306213

\bibitem[Meier, Koide \& Uchida 2001]{mei01}Meier, D., Koide, S., Uchida, Y. 2001, Science, 291, 84

\bibitem[Merloni, Heinz, \& di Matteo 2003]{mer03}Merloni A., Heinz S., di Matteo T., 2003, MNRAS, 345, 1057

\bibitem[Migliari et al. 2004]{mig04} Migliari, S., Fender, R. P., Rupen, M., Wachter, S., Jonker, P.G., Homan, J., \& van der Klis, M. 
	2004, \mnras, 351, 186

\bibitem[Miller et al. 2001]{mil01} Miller, J.~M., et al.\ 2001, \apj, 563, 928 

\bibitem[Miller et al. 2002]{mil02} Miller, J.~M., et al.\ 2002, \apjl, 570, L69 

\bibitem[Miller et al. 2004]{mil04} Miller, J.~M., et al.\ 2004, \apj, 601, 450 

\bibitem[Mirabel \& Rodr\' \i guez 1994]{mir94} Mirabel, I.~F.~\& Rodr\' \i guez, L.~F.\ 1994, \nat, 371, 46

\bibitem[Orosz et al. 2004]{oro04} Orosz, J.~A., McClintock, J.~E., Remillard, R.~A., \& Corbel, S., 2004, ApJ, submitted

\bibitem[Reig, Belloni \& van der Klis 2003]{rei03}Reig, P., Belloni, T., \& van der Klis, M.\ 2003, A\&A, 412, 229

\bibitem[Remillard 2001]{rem01} Remillard, R.\ 2001, \iaucirc, 7707

\bibitem[Revnivtsev \& Sunyaev 2001]{rev01} Revnivtsev, M.~\& Sunyaev, R.\ 2001, \iaucirc, 7715

\bibitem[Rodriguez, Corbel, \& Tomsick 2003]{rod03}Rodriguez, J., Corbel, S., \& Tomsick, J.~A.\ 2003, \apj, 595, 1032

\bibitem[Rodriguez et al. 2004]{rod04} Rodriguez, J., Corbel, S.,  Kalemci, E., Tomsick, J. A., \& Tagger, M.  2004, \apj, in press,
	 astro-ph/0405398

\bibitem[Rossi et al. 2003]{ros03}Rossi, S., Homan, J., Miller, J.~M., Belloni, T., 2003,   Proc. "The Restless High-Energy Universe" 
	(Amsterdam, May 5-8, 2003), E.P.J. van den Heuvel, J.J.M. in 't Zand, and R.A.M.J. Wijers Eds, astro-ph/0309129

\bibitem[Sanchez-Fernandez et al. 2002]{san02} Sanchez-Fernandez, C., Zurita, C., Casares, J., Castro-Tirado, A.~J., Bond, 
I., Brandt, S., \& Lund, N.\ 2002, \iaucirc, 7989

\bibitem[Stirling et al. 2001]{sti01} Stirling, A.~M., Spencer, R.~E., de la Force, C.~J., Garrett, M.~A., 
	Fender, R.~P., \& Ogley, R.~N. \ 2001, \mnras, 327, 1273

\bibitem[Sault \& Killeen 1998]{sau98} Sault R.J. \& Killeen N.E.B. 1998, The Miriad User's Guide,
        Sydney: Australia Telescope National Facility

\bibitem[Tigelaar et al. 2004]{tig04} Tigelaar, S. P., Fender, R. P., Tilamus, R. P. J., Gallo, E., \& Pooley, G. 2004,
	\mnras, in press, astro-ph/0405141


\bibitem[Tomsick, Corbel, \& Kaaret 2001]{tom01} Tomsick, J.~A., Corbel, S., \& Kaaret, P.\ 2001, \apj, 563, 229 


\bibitem[Tomsick et al. 2003a]{tom03a} Tomsick, J.~A., Corbel, S., Fender, R., Miller, J.~M., Orosz, J.~A., Tzioumis, 
	T., Wijnands, R., \& Kaaret, P.\ 2003a, \apj, 582, 933 

\bibitem[Tomsick et al. 2003b]{tom03b} Tomsick, J.~A., Kalemci, E., Corbel, S., \& Kaaret, P.\ 2003b, \apj, 592, 1100

\bibitem[Tomsick, Kalemci, \& Kaaret 2004]{tom04} Tomsick, J.~A., Kalemci, E., \& Kaaret, P.\ 2004, \apj, 601, 439 

\bibitem[Wijnands, Miller, \& Lewin 2001]{wij01} Wijnands, R., Miller, J.~M., \& Lewin, W.~H.~G.\ 2001, \iaucirc, 7715



\end{thebibliography}
\end{document}